\documentclass[%
reprint,
amsmath,amssymb,aps,
floatfix
]{revtex4}

\usepackage[utf8x]{inputenc}
\usepackage{graphicx}
\usepackage{dcolumn}
\usepackage{bm}
\usepackage{exscale}
\usepackage{tikz}
\usepackage{subfigure}

\usetikzlibrary{shadows,arrows}
\usetikzlibrary{automata}
\usetikzlibrary{positioning}

\pgfdeclarelayer{foreground}
\pgfsetlayers{main,foreground}
 
\tikzstyle{materia}=[draw, fill=white, text width=6.0em, text centered, minimum height=1.5em,drop shadow]
\tikzstyle{practica} = [materia, text width=6.8cm, minimum width=7cm,  minimum height=3em, drop shadow]
\tikzstyle{texto} = [above, text width=6em, text centered]
\tikzstyle{line} = [draw, thin, color=black, -latex']
 
\newcommand{\practica}[2]{node (p#1) [practica]  {\textbf{STEP #1}\\{\scriptsize{#2}}}}
\newcommand{\ci}{i}

\begin{document}


\title{Energy eigenfunctions of the 1D Gross--Pitaevskii equation}

\author{\v Zelimir Marojevi\'c}
\author{Ertan G\"okl\"u}
\author{Claus L\"ammerzahl}
\affiliation{ ZARM Universit\"at Bremen, Am Fallturm, 28359 Bremen, Germany}

\date{\today}   
\pacs{Valid PACS appear here}

\begin{abstract}
We developed a new and powerful algorithm by which numerical solutions for excited states in a gravito optical surface trap have been obtained. They represent solutions in the regime of strong nonlinearities of the Gross--Pitaevskii equation. In this context we also shortly review several approaches which allow, in principle, for calculating excited state solutions. It turns out that without modifications these are not applicable to strongly nonlinear Gross--Pitaevskii equations. The importance of studying excited states of Bose--Einstein condensates is also underlined by a recent experiment of B\"ucker et al in which vibrational state inversion of a Bose-Einstein condensate has been achieved by transferring the entire population of the condensate to the first excited state. Here, we focus on demonstrating the applicability of our algorithm for three different potentials by means of numerical results for the energy eigenstates and eigenvalues of the 1D Grosss--Pitaevskii--equation. We compare the numerically found solutions and find out that they completely agree with the case of known analytical solutions.
\end{abstract}

\keywords{Gross--Pitaevskii equation, stationary solutions, gravito-optical surface trap}

\maketitle


\section{Introduction}
One of the most interesting problems in today's physics is the exploration of the quantum--gravity regime. This is due to the fact that General Relativity and quantum theory are not compatible which makes it necessary to search for a new theory called quantum gravity which at the end should lead to effective modifications of General Relativity and/or quantum theory. Another issue is that in some approaches gravity is regarded as a solution to the measurement problem in quantum theory. Therefore there are a lot of reasons showing that it is important to explore the interaction of quantum matter with gravity with better accuracy. One possibility to study the behavior of quantum matter in gravitational fields is neutron and atom interferometry \cite{COW75,KasevichChu91,KasevichChu92}. One may even go further and investigate the energy eigenstates of quantum matter in a gravitational trap as has been pushed forward using ultracold neutrons at the ILL \cite{Nesvizhevskyetal00}. In this experiment the various 
eigenstates manifest themselves through a neutron flux which depends on the height in a step--like form. One difficulty in this experiment is that the steps are of the order of $\mu$m which comes from the strength of the gravitational acceleration. With the recently developed technology of Bose-Einstein condensates (BEC) in microgravity condition \cite{vanZoestetal10} another physical system is available for investigating the quantum--gravity regime for a wider range of parameters. Is is feasible to perform similar experiments with ultracold atoms in a Gravito--Optical Surface Trap (GOST) with a small and variable gravitational acceleration so that the density profile of the quantum states related to various energy levels can be measured with better resolution. 

The solution of the eingenvalue problem for the Schr\"odinger equation in such a GOST has been solved in terms of the Airy--functions in, e.g., \cite{LandauLifshitz87,springerlink:10.1007/BF00325387,PhysRevD.81.065019}. In order to be able to describe also the eigenstates for a BEC, we are solving here the eigenvalue problem for the nonlinear Schr\"odinger equations, that is, for the Gross--Pitaevskii equation (GPE). For doing so we developed in this paper a new numerical algorithm which is capable to find solutions of the GPE which belong to saddle points of the action. This algorithm is applied to three different potentials, the box, the harmonic trap and the GOST. Our numerical solutions which, among others, correspond to excited states completely agree with the known analytic solutions for the box. For a first description of the method and in order to present first results we restrict at the moment to one--dimensional problems.

Concerning a further physical motivation to study excited states it has to be mentioned that recently B\"ucker and coworkers \cite{Buecker-nature, Buecker-arxiv} demonstrated the vibrational state inversion of a Bose-Einstein condensate. This system is confined in an anharmonic trapping potential and the inversion can be achieved by controlled displacement of the trap center. By means of this procedure they transferred BECs to the first antisymmetric stationary state which is, in fact, an excited state.       

In this paper in section \ref{sec:model} we first state the problem and introduce the notation. In section \ref{sec:algorithm} we describe the newly  developed algorithm and apply this method in section \ref{sec:solutions} for solving the energy eigenvalue problem of the GPE for three different physically relevant potentials. The paper closes in Section V with an outlook indicating further work in this direction.

\section{The model}\label{sec:model}

We start with the time--dependent GPE which describes the dynamics of a BEC subject to two-particle interactions, given by the nonlinear term $g_S\vert\Psi(\mathbf{x},t)\vert^2$, and to an external potential $V_{\rm ext}$
\begin{equation}
 \ci\hbar\partial_t \Psi(\mathbf{x},t) = \left(-\dfrac{\hbar^2}{2m} \Delta + V_{\rm ext}(\mathbf{x}, t) + g_S\vert\Psi(\mathbf{x},t)\vert^2 \right) \Psi(\mathbf{x},t) \, , \label{sec_model:tdgp}
\end{equation}
where $\Psi(\mathbf{x})$, $\mathbf{x}=(x,y,z)$ is normalized to the total number of particles $N=\int_{\Omega}  |\Psi(\mathbf{x})|^2 d^3x$. The GPE is valid for dilute condensates obeying the diluteness criterion, that is, the s--wave scattering length $a$ and the average density of the gas $\bar{n}$ must fulfill $\bar{n}|a|^3 \ll 1$. The nonlinearity parameter $g_S$ is determined by the scattering length via $g_S=\frac{4\pi \hbar^2 a}{m}$, where $m$ is the mass of the atom. Moreover, the scattering length can acquire both signs, having magnitudes of some nanometers. However, in this work we will focus on the case $g_S > 0$ which describes  repulsive two--particle interactions.  The function $\Psi(\mathbf{x},t)$ has the meaning of an order parameter, is a classical field and is also interpreted as the wave function of the condensate. 

For the calculation of the ground states and higher modes of a BEC in a time--independent external potential one makes the ansatz $\Psi(\mathbf{x},t)=\Psi(\mathbf{x})\exp{(-i\mu t/\hbar)}$ leading to the stationary GPE  
\begin{equation}
\mu \Psi(\mathbf{x}) = \left(-\dfrac{\hbar^2}{2m} \Delta + V_{\rm ext}(\mathbf{x}) + g_S\vert\Psi(\mathbf{x})\vert^2 \right) \Psi(\mathbf{x})\, , \label{sec_model:tigpe}
\end{equation}
where $\mu$ is the chemical potential. We also assume that the potential $V_{\rm ext}(\mathbf{x})$ is bounded from below so that we can take $V_{\rm ext}(\mathbf{x}) \geq 0$. 

The stationary GPE can be derived from the action
\begin{equation}
A[\Psi; \mu] := F[\Psi] - \tfrac{1}{2} \mu N[\Psi] \, , \label{sec:alg:lagrangian}
\end{equation}
with the free energy 
\begin{equation}
F[\Psi] := \int_{\Omega} \left(\frac{\hbar^2}{2m} \left( \nabla\Psi(\mathbf{x}) \right)^2 + \dfrac{1}{2} V_{\rm ext}(\mathbf{x}) \Psi^2(\mathbf{x}) + \dfrac{g_S}{4} \Psi^4(\mathbf{x}) \right) \, d^3\mathbf{x}  \, ,\label{sec:alg:F}
\end{equation}
where we assumed a real $\Psi$. The particle number is given by 
\begin{equation}
N[\Psi] := \int_{\Omega} \Psi^2 \, d^3\mathbf{x} \, \label{sec:model:N}.
\end{equation}

Throughout this paper we will restrict ourselves to one--dimensional problems in order to demonstrate our new algorithm. 

In order to facilitate the numerical calculations, as one usually does, we rescale and renormalize the coordinates and the wave function according to 
\begin{equation}
x \rightarrow Lx \, , \qquad \Psi \rightarrow \sqrt{N} \Psi / L^{3/2} \, , \label{rescaling}
\end{equation}
where $\Psi(x)$ is normalized to 1, leading to 
\begin{equation}
\left(- \frac{d^2}{dx^2} + \tilde{V}_{\rm ext}(x) + \gamma \Psi^2(x) \right) \Psi(x)= \varepsilon \Psi(x), \label{sec:model:dimless_ode} 
\end{equation}
with the dimensionless quantities
\begin{equation}
\tilde{V}_{\rm ext}(x) := \dfrac{2 m L^2 V_{\rm ext}(x)}{\hbar^2}, \quad \gamma := \dfrac{2 N m g_S}{L \hbar^2}, \quad  \varepsilon := \dfrac{2 m \mu L^2}{\hbar^2}.
\end{equation}
The length scale $L$ is arbitrary and may depend on various physical parameters. It is chosen in such a way that the dimensionless quantities are convenient for numerical calculations. Note that in particular the nonlinearity parameter $\gamma$ depends on the length scale $L$. 

Note that we do not restrict ourselves to functions that are normalized to one. Instead we are searching for solutions that are not normalized for a given pair $\gamma,\varepsilon$. From equation (\ref{sec:model:dimless_ode}) it is evident that each found solution can be normalized to one by adjusting the nonlinearity $\gamma$.

\section{The algorithm}\label{sec:algorithm}

\subsection{The general setting}

\begin{figure}[t]
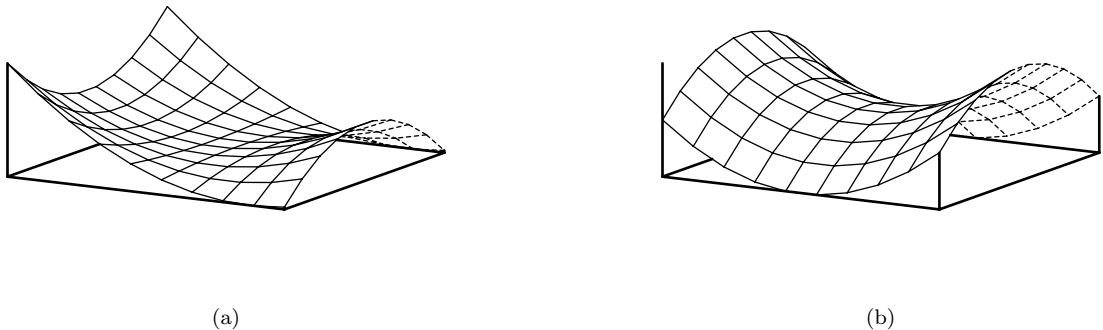

\subfigure[]{\input{monkey_saddle.tex}\label{sec:alg:fig:monkey}}
\subfigure[]{\input{horse_saddle.tex}\label{sec:alg:fig:horse}}
\caption{(a) Sketch of a monkey saddle, (b) Sketch of a horse saddle.}
\end{figure}

In computer numerics an attempt to solve nonlinear partial differential equations is to use some variant of the Newton method or the imaginary time propagation. The latter method is based on the the splitting and discretisation (e.g. Crank-Nicolson) of the unitary time evolution operator \cite{Muruganandam20091888,Vudragovic20122021}. This is reliable for ground state solutions. In this paper we will present a new Newton Method.

Newton methods are gradient based algorithms that follow a descent direction until a local minimum of the action is reached. The direction depends on the choice of the inner product and/or of the preconditioning procedure. Solutions can be understood as critical points of some action $A$ on the underlying dual space defined by the inner product. The type of a critical point related to a solution is determined by the eigenvalues of the Hessian \footnote{ By the Hessian we denote the operator $\mathcal{H}[u]$ which appears within the Taylor expansion of some action $A[u+\alpha h] = A[u] + \alpha \langle \nabla_{L_2}\mathcal{A}[u], h\rangle + \dfrac{1}{2} \alpha^2 \langle \mathcal{H}[u] h, h\rangle + \mathcal{O}(\alpha^3)$. In a finite difference approximation this is represented by a finite dimensional matrix.} evaluated at the critical point: 
\begin{itemize}\itemsep=-2pt
\item If all eigenvalues of the Hessian are positive then the critical point is a local minimum of $A$. 
\item On the other hand if all are negative then we have a local maximum. 
\item If we have a finite number of negative eigenvalues and all other eigenvalues are $>0$ then we have a horse saddle Fig. \ref{sec:alg:fig:horse}. In this situation the number of negative eigenvalues is the number of linear independent descent directions at this critical point. 
\item And the last case is when the Hessian is degenerated at a critical point. For example this can be associated with a monkey saddle Fig. \ref{sec:alg:fig:monkey} for isolated critical points. 
\end{itemize}

The number of negative eigenvalues is known as the Morse index. Solutions that belong to a local minimum of an action $A$ are candidates for solutions which can be easily found by standard Newton methods, which searches in the whole $L_2$ space. Unfortunately, finding critical points of a certain saddle type depends on an educated guess. In order to have a straightforward method at hand it is necessary to confine the search on a subspace of our Hilbert space. For linear eigenvalue problems this is easy to do because of the orthogonality of eigenfunctions. The gradient at every iteration step is orthogonalized with respect to the previously found eigenfunctions using the Gram-Schmidt procedure. Therefore it is easy to find eigenfunctions in ascending order of eigenvalues or, equivalently, in ascending order of the Morse index. Unfortunately, in the nonlinear case the orthogonality no longer holds, so that an other approach is needed. The basic idea is to constrain the quest for a solution to a submanifold in 
the underlying function space.

In the following we need the first variational derivative (G\^ateaux derivative) 
\begin{equation}
A^{\prime}[\Psi;\mu] h := \dfrac{d}{d\epsilon} A[\Psi+\epsilon h; \mu] \bigg|_{\epsilon=0} \, , \label{sec:alg:def:gateaux}
\end{equation}
which via
\begin{equation}
\langle \nabla_{L_2} \mathcal{A}[\Psi;\mu], h \rangle := A^{\prime}[\Psi;\mu] h \, ,
\end{equation}
can be identified with an $L_2$ gradient $\nabla_{L_2} \mathcal{A}[\phi^k,\mu]$. Here $\langle \cdot , \cdot \rangle$ denotes the $L_2$ scalar product. For the GPE we have
\begin{equation}
\nabla_{L_2} \mathcal{A}[\Psi;\mu] = -\frac{d^2}{dx^2} \Psi + (V_{\rm ext} - \mu) \Psi + \gamma \Psi^3 \, . \label{sec:alg:L2grad}
\end{equation}
Therefore, if $\Psi$ is a critical point of the action $A$ then the $L_2$ gradient of $\mathcal{A}$ vanishes and, hence, $\Psi$ is a solution of the GPE. 

\subsection{Review of the Newton method}

The discrete Newton method is given by
\begin{equation}
\phi^{k+1} = \phi^k - \tau d^k \, \label{sec:alg:newton_method},
\end{equation}
where 
\begin{equation}
d^k :=\mathcal{O}^{-1}  \nabla_{L_2} \mathcal{A}[\phi^k;\mu]
\end{equation}
is the search direction and $\mathcal{O}^{-1}$ denotes a preconditioning operator that improves the convergence behaviour. 
In this context $k$ is the iteration index and $\tau$ the stepsize. The minus sign in front of $\tau$ denotes that the correction of the step $\phi^k$ is performed in the negative direction of the preconditioned $L_2$ gradient. The stepsize can be a constant or can be determined at every iteration step using the linesearch or trusted region method. The solution then is given by $\phi^{\rm sol} := \lim_{k \rightarrow \infty} \phi^k$. In the nonlinear case the widely used Newton method is only capable to find Morse index zero solutions. Finding higher Morse index solutions for strong nonlinearities is a hard task and the standard Newton method is not able to do that. 

A von Neumann analysis applied to the standard Newton method (i.e. with the preconditioning $\mathcal{O}^{-1} = 1$ ) leads to a convergence criterion like the famous Courant-Friedrich-Lewy condition which estimates a bound on the stepsize $\tau$ depending on the discretization lengths and other parameters of the differential equation. Therefore a bad choice for $\tau$ causes a failure of the Newton method. A too small $\tau$ decreases the convergence rate. In order to handle this issue the preconditioning $\mathcal{O}^{-1}$ is necessary. There are two well known methods, among others:
\begin{enumerate}\itemsep=-2pt
\item A classical choice for $\mathcal{O}^{-1}$ is the inverse of the Hessian, or at least a numerical approximation. Due to the fact that computation time and storage space are precious and the full inverse Hessian is a dense matrix that is not fast computable new techniques have been invented to overcome this problem. The simplest one is to use the difference between two $L_2$ gradients  approximating the diagonal of the Hessian. 
\item A modern approach is to use the Sobolev preconditioning \cite{Neuberger200912,Richardson2000241}. The $L_2$ gradient is mapped to a different Sobolev space, for example $W^{1,2}$. From a mathematical point of view the $L_2$ gradient is filtered in Fourier space so that spatial oscillations are smoothed out. 
\end{enumerate}
Upon the choice of preconditioning the direction of the gradient is altered. For a descend direction we have $\langle \mathcal{O}^{-1} \nabla_{L_2} \mathcal{A}, \nabla_{L_2} \mathcal{A} \rangle > 0$ for some arbitrary action $A$. The stepsize control of classical Newton methods fails if this condition does not hold. 

\subsection{Newer approaches}

From equation (\ref{sec:alg:F}) it is evident that $F[\phi^k]>0$ for any $\phi^k \neq 0$ and $g_S>0$ so that $\phi^k=0$ is the only critical point of $F$. Therefore it is only the term $\mu N$ in  (\ref{sec:alg:lagrangian}) by which new critical points can appear. Accordingly, the key idea for the existence of solutions of non linear differential equations is to have terms that are capable of balancing the non linearity and all other terms. 

In order to emphasize the idea of balancing the nonlinearities we consider for demonstration purpose a classical case \cite{Choi1993417,Horak2004} for the situation without external potential and attractive two-particle interaction, thus $\tilde{V}_{\rm ext}=0$ and $\gamma<0$. Then the functional $F$ reads
\begin{equation}
F[\phi^k] = \int_{\Omega} \left(\frac{1}{2} \left(\frac{d}{dx}\phi^k \right)^2 + \frac{1}{4} \gamma \left(\phi^k \right)^4 \right) dx \,\label{sec:algorithm:ex1} . 
\end{equation}

$F[0]=0$ is a critical point and $F$ is not bounded. Without further constraints a standard Newton method would fail. It is clear that there exists a $t^k \neq 0$ that fulfils
\begin{equation}
F^{\prime}[t^k \phi^k] \phi^k = \int_{\Omega} \left( \frac{d}{dx}\phi^k \right)^2 dx + \left( t^k \right)^2 \gamma \int_{\Omega} \left( \phi^k \right)^4 dx = 0 \, , \label{sec:alg:school1}
\end{equation}
which means that the kinetic part is balancing the interaction part. 

A Newton method which calculates the L2 gradient at the point $t^k\phi^k$ defined by equation (\ref{sec:alg:school1}) generates a sequence $\lbrace t^k,\phi^k \rbrace$ where the $\phi^k$ converge to a solution $\phi^{\rm sol} \neq 0$ in $L_2$, which is a local minimum of $F$. This is known as a minimization process restricted to the Nehari manifold
\begin{equation}
t_{\rm ref} = \text{extremum} \, F[t^k \phi^k] \, . \label{sec:alg:nehari_manifold}
\end{equation}

For the $k$-th step, $t_{\rm ref}^k \phi^k$ is a reference point in the underlying function space where the $L_2$ gradient is calculated. The definition (\ref{sec:alg:nehari_manifold}) is equivalent to $F^{\prime}[t^k_{\rm ref}\phi^k] \phi^k = 0$. The sequence of functions $\phi^k$ calculated this way will converge to the solution $\phi^{\rm sol}$. With the restriction to the Nehari manifold it is possible to find Morse index one solutions of (\ref{sec:algorithm:ex1}). (For a general functional this may not always work.) Therefore it is, in general, very useful to confine the search for solutions to a manifold where the critical point lies in a local minimum on this manifold so that classical Newton methods are able to find such solutions. If one finds no extremum then the method is not applicable.  

A generalization of this idea has been presented in \cite{yao:937}. There, in the $k$-th iteration step the function
\begin{equation}
P^{D, k}_{\vec t} := \sum_{i=1}^{D} t_i^k \Upsilon_i + t^k \phi^k \, ,
\end{equation}
has been defined where $\vec{t} := \left(t_1^k \dots t_D^k, t^k\right)$ and the $\Upsilon_i$ are the previously found solutions, which were calculated by the algorithm \cite{yao:937}, and $D$ is the dimension of the support that is spanned by the $\Upsilon_i$. 

The idea behind this is to find solutions in the order of their Morse index which is similar to linear problems. First find the global minimum, then use the ground state to define a solution manifold in order to stay away from the ground state. After the the first excited state is found, it is used again together with the ground state to define a new solution manifold in order to stay away from the first and second solution. This is repeated until the algorithm fails. The ground state has Morse index zero and the first excited state has Morse index one. 

Then, for some action $J [P^{D,k}_{\vec t}]$ what can be interpreted as function of $\vec{t}$, the extrema of the $J [P^{D,k}_{\vec t}]$ determine the vector ${\vec t}$ which is taken to define the reference point  
\begin{equation}
{\vec t}_{\rm ref } = \left(t_1^k \dots t_D^k,  t^k\right) = \text{extremum} \, J [P^{D,k}_{\vec t}] \label{sec:alg:minmax_ref} \, .
\end{equation}
Then the $P^{D,k}_{\vec t_{\rm ref}}$ is the reference point at which the $L_2$ gradient $\nabla_{L_2} \mathcal{J}[P^{D,k}_{\vec t_{\rm ref}}]$ is calculated.

If $\gamma>0$ then the action may have minima. Zhou uses in \cite{yao:937} the so called active Lagrangian $J:=F-\tfrac{1}{2} \varepsilon^k (N-1)$ for his algorithm in order to find normalized solutions. The eigenvalue term $\varepsilon^k$ is indispensable, but the final eigenvalue is not known from the beginning and has to be altered at every iteration step $k$. This includes the risk that at some iteration step $k$ the solution is the trivial one $\vec{t}_{\rm ref} = 0$. Zhou demonstrated that for $\gamma \in \mathcal{O}(1)$ in a 2D GPE with a 2D harmonic trapping it is possible to find solutions in the order of their eigenvalues $\varepsilon_i$ where $\varepsilon_i > \varepsilon_{i-1}$. 

A related way to define a solution manifold \cite{Jianxin200466} is to require that the directional derivatives in direction of the $D$ known solutions and the current iterations step vanishes. In order to find the reference point $\vec{t}_{\rm ref}$ the following system of equations has to be solved:
\begin{align}
\left< \nabla_{L2} \mathcal{J}[P^{1,k}_{\vec{t}_{\rm ref}}], \Upsilon_1 \right> &= 0 \nonumber\\
&\vdots \nonumber\\
\left< \nabla_{L2} \mathcal{J}[P^{1,k}_{\vec{t}_{\rm ref}}], \Upsilon_D \right> &= 0 \nonumber\\
\left< \nabla_{L2} \mathcal{J}[P^{1,k}_{\vec{t}_{\rm ref}}], \phi^k \right> &= 0 . \label{sec:alg:orthomin_ref}
\end{align}
This is a system of $D+1$ equations for the $D+1$ unknown variables $(t^k_1,...,t^k_D,t^k)$. The trivial solution is always a solution, but not the desired one. The numerical solution of this system depends on the initial guess of the vector $(t^k_1,...,t^k_D,t^k)$. Due to the nonlinearity of $\nabla_{L2} \mathcal{J}[P^{1,k}_{\vec{t}_{\rm ref}}]$ this system has more than one solution for suitable $(\varepsilon^k,\phi^k)$. In order to find an optimal reference point the initial vector $(t^k_1,...,t^k_D,t^k)$ has to be guessed systematically. We used all corners and all center points of the faces of a $D+1$ dimensional cube as guesses and took the initial guess where $\forall_{i \leq D} \vert t^0_i \vert < \vert t^0 \vert$. Under this condition we always succeeded to find  a new solution for nonlinearities of the order $\mathcal{O}(1)$. With increasing nonlinearity it becomes impossible to fulfil this condition when a solution with smaller non linearity is used as a guess. 
As a result of our numerical simulations, we like to add some remarks concerning the feasibility of the methods presented in \cite{yao:937} and \cite{Jianxin200466}: 
\begin{itemize}
\item First, for the reference point (\ref{sec:alg:minmax_ref}) and (\ref{sec:alg:orthomin_ref}) one encounters the problem that the old solutions also lie on the same manifold, so that if the initial guess is too far away from the final solution then one finds only old critical points. 
\item Second, due to the fact that the eigenvalue $\varepsilon^k$ is altered at every step there is a chance that it converges to an old one. 
\item Third, in higher dimensions for non isotropic potentials the order of eigenvalues and eigenfunctions depending on the parameters of $\tilde{V}_{\rm ext}$ can be changed and the guess for the next solution may not be an appropriate one.
\end{itemize}

\subsection{The new algorithm}

\begin{figure}[h]
\begin{center}
\begin{tikzpicture}[scale=1.0,transform shape,->,>=stealth']
\path \practica {1}{Calculate the Eigenfunction $\Psi_n$ for \mbox{$\gamma=0$}. \\ Calculate the Eigenvalue $E_n$ for \mbox{$\gamma=0$}. \\ Set \mbox{$s\leftarrow0, \phi^0_{n,0}\leftarrow\Psi_n, \psi_{n,0}\leftarrow\Psi_n$} \\ \mbox{$\mu_{n,0}=\lceil E_n \rceil+\Delta\mu, \gamma_{n,0} = 1$}  };

\path (p1.south)+(0.0,-1) \practica{2}{ if \mbox{$k = 0$} then find optimal ref. point \mbox{$\vec{t}^{k}_{\rm ref}$} else find the new ref point using \mbox{$\vec{t}^{k-1}_{\rm ref}$} as a guess. };

\path (p2.south)+(0.0,-1) \practica{3}{Calculate \mbox{$\nabla_{L_2}\mathcal{A}[P^{1,k}_{\vec{t}_{\rm ref}}]$} };

\path (p3.south)+(0.0,-1) \practica{4}{if \mbox{$\|\nabla_{L_2}\mathcal{A}[P^{1,k}_{\vec{t}_{\rm ref}}]\|_{\infty} < \eta $}  };

\path (p3.east)+(5.0,-1.0) \practica{7}{Calculate $N$ \\ Store \mbox{$\left(\mu_{n,s},\gamma_{n,s} N,N^{-1/2} P^{1,k}_{\vec{t}_{\rm ref}}\right)$}.  \\ \mbox{$\gamma_{n,s+1} \leftarrow \gamma_{n,s} N$}  \\ Replace \mbox{$\psi_{n,s}$} with \mbox{$P^{1,k}_{\vec{t}_{\rm ref}}$} \\ \mbox{$\mu_{n,s+1} \leftarrow \mu_{n,s} + \Delta \mu$} \\ \mbox{$s \leftarrow s+1\, , \;\; k \leftarrow 0$} \\ if \mbox{$\mu_n < \mu_{n,{\rm fin}}$}};

\path (p4.south)+(0.0,-1.6) \practica{5}{Solve: \\ $\left( -\dfrac{d^2}{dx^2} + (V_{\rm ext}-\mu_{n,s}) + 3\gamma_{n,s} (P^{1,k}_{\vec{t}_{\rm ref}})^2\right) d^k = \nabla_{L_2}\mathcal{A}[P^{1,k}_{\vec{t}_{\rm ref}}]$};

\path (p5.south)+(0.0,-1) \practica{6}{ \mbox{$\phi_{n,s}^{k+1} = \phi_{n,s}^k - \tau$sgn$(q^k_s)d^k$} \\ \mbox{$k \leftarrow k+1$} };

\path (p7.south)+(0.0,-1) \practica{8}{ Exit };

 \path [line] (p1) -- (p2);
 \path [line] (p2) -- (p3);
 \path [line] (p3) -- (p4);
 \path [line] (p5) -- (p6);

 \path (p4)  edge node[anchor=north,below,text width=4cm,yshift=0.5em,xshift=7em] {no} (p5);
 \path (p7)  edge node[anchor=north,below,text width=4cm,yshift=0.5em,xshift=2.3cm] {yes} (p8);
 \path (p7.north)  edge[bend right=15] node[anchor=east,below,text width=4cm,yshift=1.5em,xshift=4em] {no} (p2.east);
 \path (p4.east)  edge node[anchor=north,below,text width=4cm,yshift=1.5em,xshift=5em] {yes} (p7);
 \path (p6.west)  edge[bend left=20] node {} (p2.west);
\end{tikzpicture}
\end{center}
\caption{Flow chart of the algorithm} \label{sec:alg:flow_chart}
\end{figure}
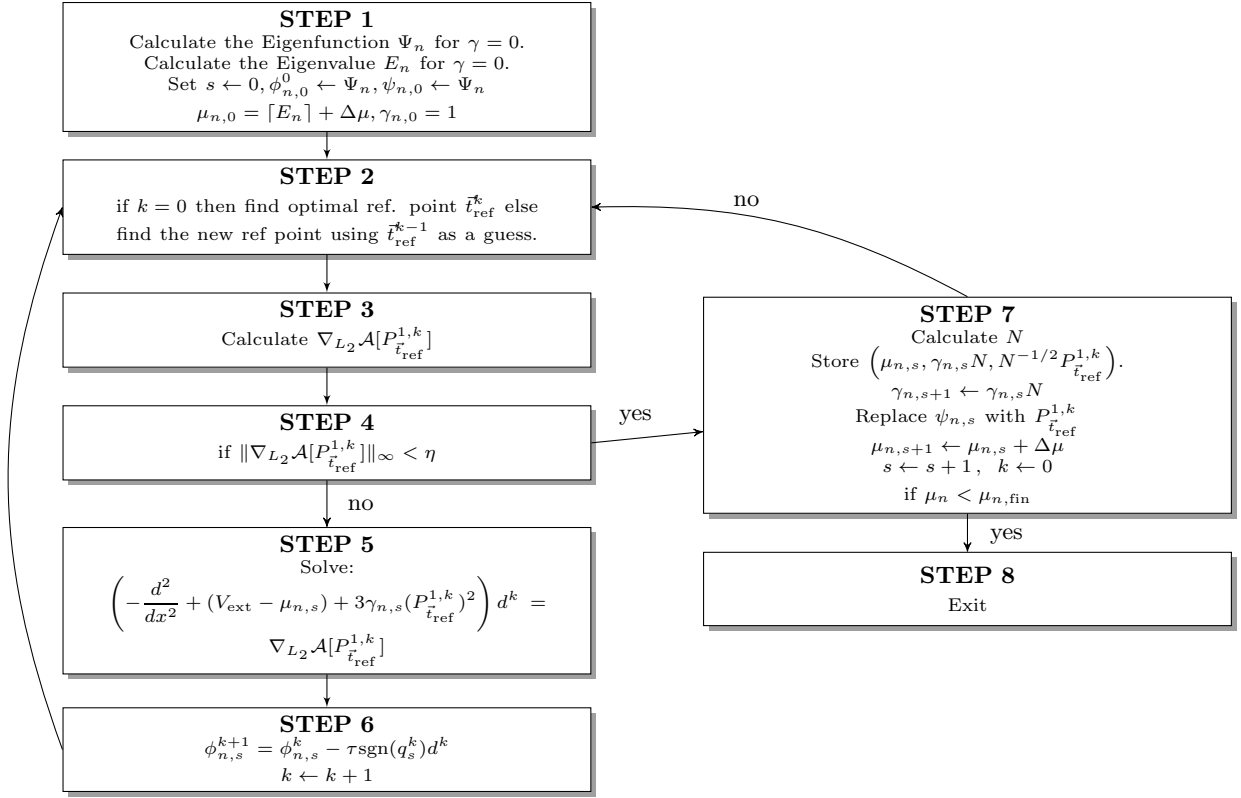

In order to overcome these problems, we developed a modified algorithm (see Fig. (\ref{sec:alg:flow_chart})) which yields the following characteristics:
\begin{enumerate}
\item Instead of using the previously found solutions $\Psi_1\cdots \Psi_D$, we are calculating the reference point via 
\begin{align}
\left< \nabla_{L2} \mathcal{A} \left[ r_s^k \psi_{n,s} + q_s^k\phi_{n,s}^k ; \mu_{n,s} \right], \psi_{n,s} \right> &= 0 \nonumber\\
\left< \nabla_{L2} \mathcal{A} \left[ r_s^k \psi_{n,s} + q_s^k\phi_{n,s}^k ; \mu_{n,s} \right], \phi_{n,s}^k \right> &= 0 \label{sec:alg:myrefpoint} .
\end{align}
That means that we do not need the previous $D$ solutions of lower Morse index as in (\ref{sec:alg:orthomin_ref}).  
Here $r_s^k, q_s^k \in \mathbb{R}$ and $k$ and $s$ are numerical counter variables which are only used in the algorithm.
The quantum number $n$ refers to the mode of the solutions we are interested in and is defined through the linear eigenfunctions.
The functions $\psi_{n,s}$ and $\phi_{n,s}^k$ have the same nodal structure and $\phi_{n,s}^k$ can be viewed as a correction to the solution $\psi_{n,s}$ for the previous eigenvalue.

\item Unlike in the previously presented approaches we are working now with a fixed eigenvalue $\mu_{n,s}$ which is increased by a value $\Delta \mu$ after a solution is found for the current $\mu_{n,s}$, thus $\mu_{n,s+1} = \mu_{n,s} + \Delta\mu$. The nonlinearity $\gamma_{n,s}$ is determined as a function of this  chosen $\mu_{n,s}$. 
\item The solution found in this way is not normalized to one.  Therefore we normalize it and readjust the $\gamma_{n,s}$ according to the particle number $N$.
\item  For the search direction $d^k$, 
\begin{equation}
d^k = \mathcal{O}^{-1} \nabla_{L_2} \mathcal{A} \left[ \phi_{n,s}^k; \mu_{n,s} \right] = \left( -\frac{d^2}{dx^2} + \left( V_{\rm ext} - \mu_{n,s} \right) + 3 \gamma_{n,s} \left(\phi_{n,s}^k\right)^2 \right)^{-1} \nabla_{L_2} \mathcal{A} \left[ \phi_{n,s}^k; \mu_{n,s} \right] \, , \label{sec:alg:myprecond}
\end{equation}
we are using the inverse of the analytic Hessian evaluated at $\phi_{n,s}^k$ as the preconditioning operator $\mathcal{O}^{-1}$.
Our reference point defined by (\ref{sec:alg:myrefpoint}) together with the preconditioning (\ref{sec:alg:myprecond}) assures that we do not leave the subspace with same nodal structure. 
In contrast, a Sobolev preconditioning would lead to ground state solutions only. 
\end{enumerate}
  
The algorithm is implemented in C++. The main part consists of two nested loops with the inner loop counter $k$ and the outer loop counter $s$. The inner loop (STEP 1 to STEP 6) represents our Newton method. Within the outer loop $\mu_{n,s}$ is increased and storage operations are conducted. For convenience we take the calculated $\gamma_{n,s}$ as starting point for the solution for the next eigenvalue $\mu_{n,s+1}$. 

For the numerical derivatives a three point stencil is used. The integrals are evaluated with Simpson's rule and the differential equation from STEP 5 is solved in a finite difference setup with a Bi Conjugate Gradient solver. We used $\Delta \mu = 0.5$ , $\tau=0.01$, ${\rm dx}=0.05$ and $1401$ grid points.

In the following we present the individual computational steps as depicted in Fig. (\ref{sec:alg:flow_chart}). 

\begin{description}
\item[STEP 1] The functions $\psi_0$ and $\phi_{n,0}^0$ are initialized to $\Psi_n$, where $\Psi_n$ is the analytic eigenfunction of the linear Schr\"odinger equation for the $n$-th quantum number. However, if ground state solutions are to be calculated, then one has to set $\psi_n\leftarrow0$. In this case the solution manifold reduces to the Nehari manifold. At the end, set $\gamma_{n,s}\leftarrow1$.

\item[STEP 2] The numerical algorithm which solves the system of the two equations (\ref{sec:alg:myrefpoint}) needs an initial guess for $\vec{t}^k_{\rm ref}$. Due to the nonlinearity of $\nabla_{L_2} \mathcal{A}\left[\phi_{n,s}^k; \mu_{n,s} \right]$ with respect to the $\phi_{n,s}^k$ the solution is not unique. As guesses we used $\vec{t}^0_{\rm ref} = (0,1)$ and $\vec{t}^0_{\rm ref} = (1,1)$ and selected for convenience the final $\vec{t}^0_{\rm ref}$ with the larger vector norm.
\item[STEP 3] Calculate the $L_2$ gradient using (\ref{sec:alg:L2grad}).
\item[STEP 4] Test for the quality of convergence. We use the maximum norm to check for convergence (the $L_2$ norm could be used also since both norms are equivalent). For our numerical calculations we set  $\eta=1\cdot 10^{-5}$. On expense of more iteration steps better results can be achieved for smaller $\eta$.
\item[STEP 5] In general, the numerical inversion of the operator $\mathcal{O}$ on the l.h.s. of (\ref{sec:alg:myprecond}) consumes much computer storage and calculation time. In order to avoid this, we solve the differential equation of STEP 5 in figure \ref{sec:alg:flow_chart} and, thus,  obtain the search direction $d$. The disadvantage of this procedure is that the differential operator $\mathcal{O}$ has to be assembled at every iteration step.
\item[STEP 6] A modified Newton step is carried out. We set the stepsize to a constant value $\tau=0.01$. A classical linesearch or trusted region stepsize control is not applicable due the fact that $d$ is not always a descent direction. Note that equation (\ref{sec:alg:myrefpoint}) is invariant under the simultaneous sign change of $r^k_s$ and $q^k_s$. This reflects the invariance of the GPE under the transformation of the wave function $\psi \rightarrow -\psi$. Thus, the factor $\text{sgn}(q_s^k)$ has to be introduced into the second term in the r.h.s. of (\ref{sec:alg:newton_method}) in order to have a unique notion of ascent and descent directions, respectively. Therefore, the search direction can be made unique by means of multiplying $\tau d[\phi_{n,s}^k; \mu_{n,s}]$ with $\text{sgn}(q_s^k)$. 
\item[STEP 7] First we calculate the particle number $N$ of $P^{1,k}_{\vec{t}_{\rm ref}}$ according to (\ref{sec:model:N}). Then we save the solution $P^{1,k}_{\vec{t}_{\rm ref}}$ for the current eigenvalue $\mu_{n,s}$, the adjusted nonlinearity $\gamma_s$, and the corresponding normalized solution $N^{-1/2} P^{1,k}_{\vec{t}_{\rm ref}}$. After that we replace the function $\psi_{n,s}$ by the just calculated $P^{1,k}_{\vec{t}_{\rm ref}}$. Before we proceed to the next inner loop $s+1$ we increase the eigenvalue $\mu_{n,s}$ by $\Delta\mu$, where $\Delta\mu$ has a typical value of $0.5$. Go to STEP 2.
\end{description}

\begin{figure}[h]
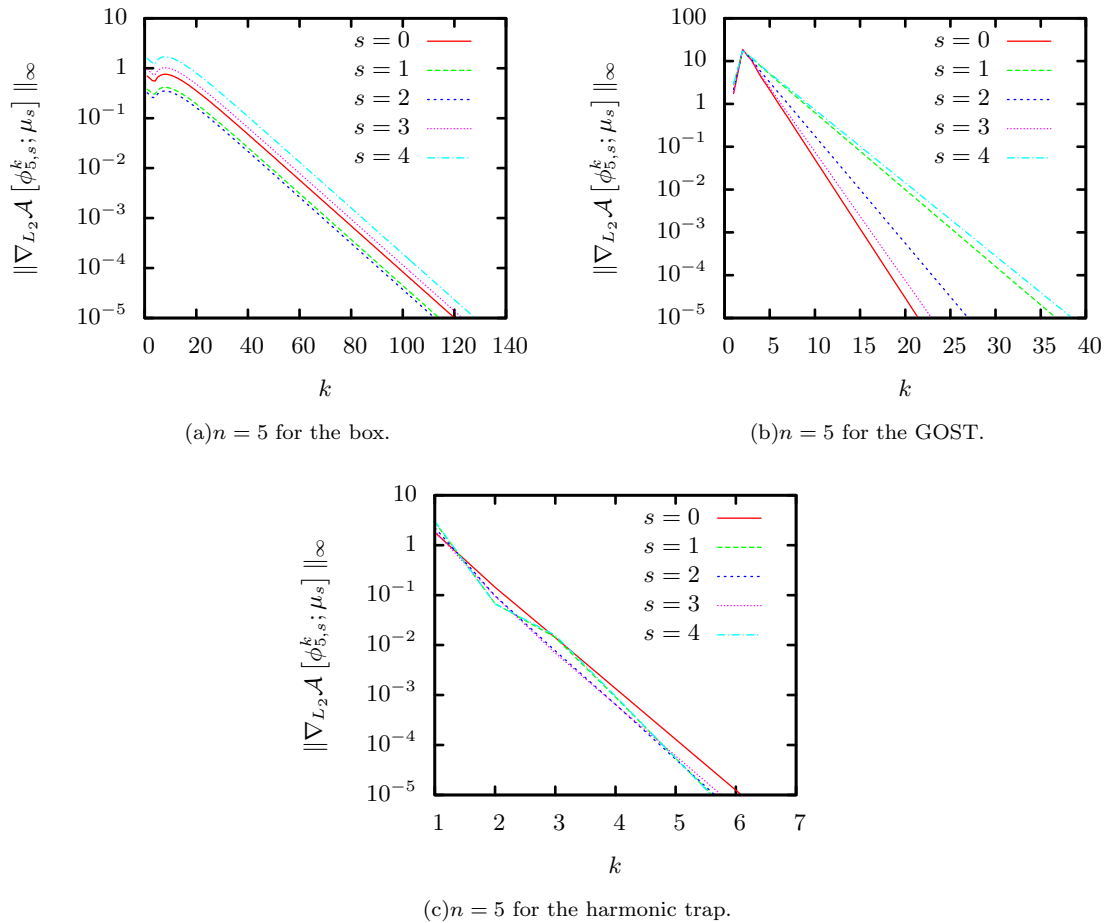

\subfigure[$n=5$ for the box.]{\input{maxnorm_von_i_box.tex}\label{sec:alg:fig:maxnorm_von_i}}
\subfigure[$n=5$ for the GOST.]{\input{maxnorm_von_i_gost.tex}\label{sec:alg:fig:maxnorm_von_i_gost}}
\subfigure[$n=5$ for the harmonic trap.]{\input{maxnorm_von_i_trap.tex}\label{sec:alg:fig:maxnorm_von_i_trap}}
\caption{Maximum norm of the $L_2$ gradient as a function of the iteration counter $k$ for the first five outer loops iterations.}
\end{figure}

In Figs.~\ref{sec:alg:fig:maxnorm_von_i}-\ref{sec:alg:fig:maxnorm_von_i_trap} we show typical forms of the error estimate $\| \nabla_{L_2} \mathcal{A} [\phi_{n,s}^k; \mu_{n,s}]\|_{\infty}$ as a function of the inner loop counter $k$ for each of the three problems discussed later. The number of iteration steps for this algorithm applied to these problems was of the order of $\mathcal{O}(10^2)-\mathcal{O}(10^3)$ for the inner loop. 
From these graphs it is evident that the error estimate is not necessarily decreasing right from the first iteration step $k=0$ as one might expect. Thus our algorithm permits that the search direction $\langle \nabla_{L_2} \mathcal{A}\left[\phi_{n,s}^k; \mu_{n,s} \right], d^k \rangle \lessgtr 0$ can be descending or ascending in contrast to the aforementioned algorithms (see Appendix \ref{sec:apd:1}). We have to emphasize that this depends on the preconditioning.  
In these logarithmic plots the linear behaviour reveals the exponential decay of the norm.

\section{Solutions for various potentials}\label{sec:solutions}

In this section we present analytical and numerical solutions for the energy eigenstates and the energy eigenvalues for the GPE for three potentials, that is, (i) for a box, (ii) gravitational surface trap, and (iii) the harmonic trap. While usually in experiments BECs are created in the ground state, excited states might emerge through an appropriate periodic motion of, e.g., the walls of a box potential. This is similar to the creation of waves of a viscous fluid in a box through the motion of walls. The explicit procedure of the creation of excited states of a BEC obeying the GPE will be discussed in a subsequent paper.

\subsection{BEC in a box}

In this section we present the numerical results of a BEC confined in a box of finite size $L$. With (\ref{rescaling}) and the natural length scale $L = \hbar/\sqrt{2 m \mu}$ the dimensionless GPE reads
\begin{equation}
\left(-\dfrac{d^2}{dx^2} + \tilde{V}_{\rm ext}(x) + \gamma \Psi_n(x)^2\right) \Psi_n(x) = \varepsilon_n \Psi_n(x) \, . \label{sec:GOST:1d_box_eq}
\end{equation}
The potential is given by
\begin{equation}
\tilde{V}_{\rm ext}(x) = 
\begin{cases}
0 &\text{if } x \in [0, 1]\\
\infty &\text{else } .
\end{cases}
\end{equation}
As usually, we require the standard boundary conditions $\Psi_n(0) = 0$ and $\Psi_n(1) = 0$. 

For $\gamma=0$ the eigenfunctions and energies are simply given by
\begin{equation}
\Psi_n(x) = \sqrt{2} \sin(\pi n x) \qquad \text{and} \qquad \varepsilon_n = \pi^2 n^2 \,  ,\end{equation}
where $n = 1, 2, 3, \ldots$.

For $\gamma>0$ this problem can be solved analytically by means of the Jacobi elliptic function sn \cite{PhysRevA.62.063610}
\begin{equation}
\Psi_n(x) = 2\sqrt{2m \gamma^{-1}} n \text{K}(m) \text{sn} \left( 2 n \text{K}(m) x \vert m \right) \label{sec:box:nonlinana} \, , 
\end{equation}
where K is the complete elliptic integral of the first kind. The definitions of the elliptic integrals and functions are taken from  \cite{abramowitz+stegun}. $2 n K(m)$ is the real period of the Jacobi sn function. The modulus $m$ of the Jacobi sn function is determined by the following equation 
\begin{equation}
8 n^2 \left( \text{K}(m)-\text{E}(m)\right) = \gamma \, ,  \label{sec:box:mod}
\end{equation}
which is derived from the normalization condition and $\text{E}(m)$ is the complete elliptic integral of the second kind. The energy spectrum then is
\begin{equation}
\varepsilon_n = 4 n^2 \text{K}(m)^2 (1+m) \, .
\end{equation}

The limiting case $\gamma=0$ leads to $m=0$, $K(0)=\tfrac{\pi}{2}$, $E(0)=\tfrac{\pi}{2}$, and the Jacobi elliptic function sn in equation (\ref{sec:box:nonlinana}) reduces to $\sin{(\pi n x)}$.

\begin{figure}[h]
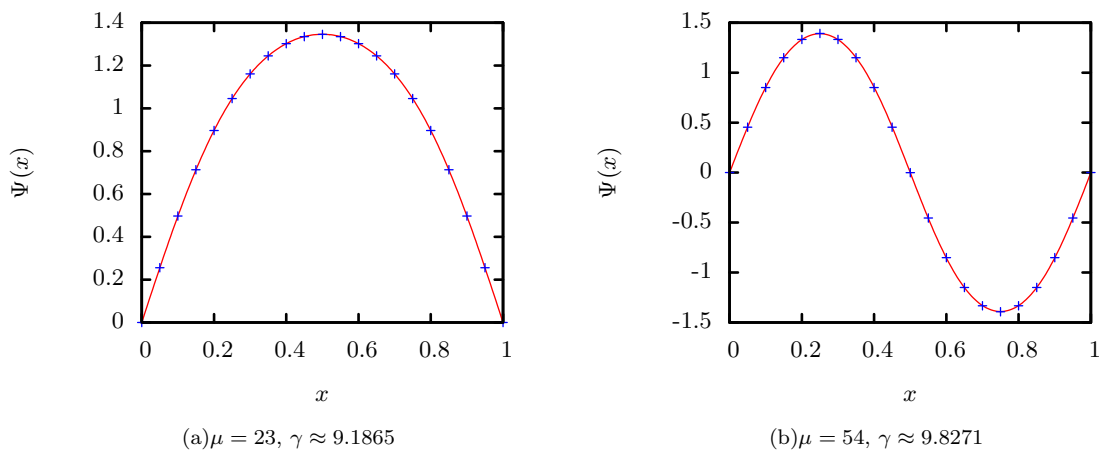

\subfigure[$\mu=23$, $\gamma\approx 9.1865$]{\input{comp_num_ana_0.tex} \label{sec:box:fig:comp_num_ana_0}}
\subfigure[$\mu=54$, $\gamma\approx 9.8271$]{\input{comp_num_ana_1.tex} \label{sec:box:fig:comp_num_ana_1}}
\caption{Comparison of numerical solution (solid red line) and the analytic solution (dots).}
\end{figure}

The knowledge of these analytically given solutions is very useful as a benchmark for our algorithm. In Figs.~\ref{sec:box:fig:comp_num_ana_0} and~\ref{sec:box:fig:comp_num_ana_1} the comparison between the analytical and numerical solution for the ground state and the first mode is shown. The solid red line is the numerical result and the dots are calculated with the analytic solution. The deviations are of the order of $\mathcal{O}(10^{-4})$. For large nonlinearities the numerical calculation of modulus with equation (\ref{sec:box:mod}) becomes difficult because the number of required decimals increases fast. 

\begin{figure}[h]
\input{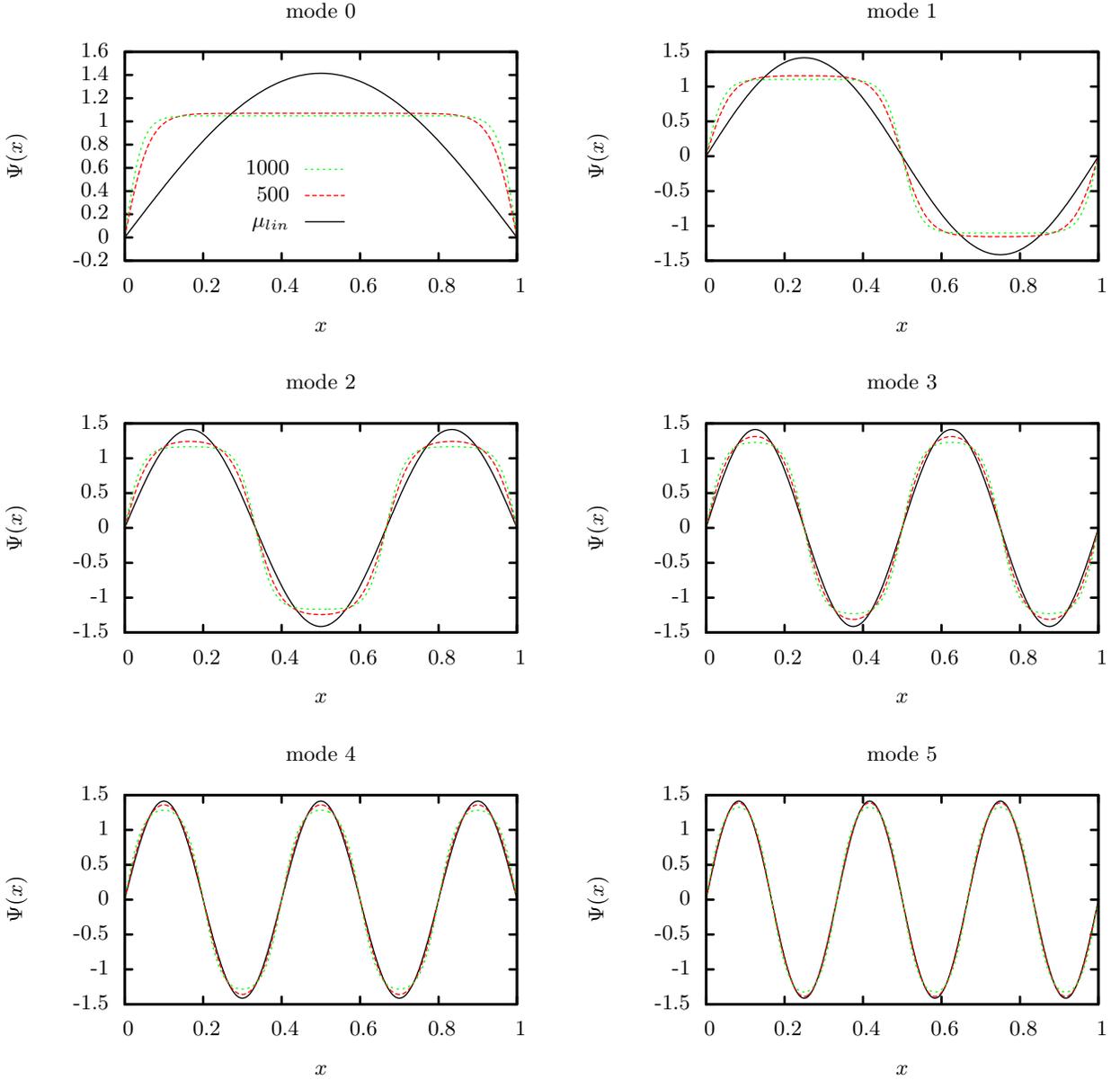}
\caption{The first six solutions of the GPE in a box. The corresponding nonlinearities $\gamma_n$ for given $\mu_n$ can be found in table (\ref{sec:box:table1}).}
\label{sec:box:fig:1d_box}
\end{figure}

In Fig.~\ref{sec:box:fig:1d_box} the first six modes are plotted, starting with the strictly positive zeroth mode. 
The solid black lines show the eigenfunctions for the linear case $\gamma = 0$. The other two lines show the numerical solutions of the nonlinear problem for different eigenvalues $\mu_{n}$. All solutions are normalized to one. The corresponding eigenvalues $\mu_{n}$ for a given $\gamma_{n}$ can be read off from table \ref{sec:box:table1} or from Fig. \ref{sec:box:fig:1d_gamma_von_mu_box}. The relation between $\gamma_{n}$ and $\mu_{n}$ in Fig.~\ref{sec:box:fig:1d_gamma_von_mu_box} is proportional but not linear.

The amplitudes of the wave function shown in Fig. \ref{sec:box:fig:1d_box} decrease with increasing $\gamma_n$ and the maxima and minima become more and more flat. This can be easily understood from the fact that the repulsion becomes stronger for larger $\gamma_n$ so that the wave functions tend to a spatial equalization. As a consequence, the gradient of the wave functions at the boundary grows and with that the kinetic energy. 

Note that for increasing mode numbers and at fixed nonlinearity $\gamma$ the broadening effect gets smaller. This can be seen from the solutions  
for $\mu_0=500$ at mode zero and for the 5-th mode at $\mu_5=1000$ (see Fig. \ref{sec:box:fig:1d_box}). 

\begin{figure}[h]
\input{1d_gamma_von_mu_box.tex}
\caption{The one-to-one correspondence between the energy eigenvalues $\mu$ and the nonlinearity parameter $\gamma$ for the box potential for different eigenmodes. The solid line represents the ground state, increasing mode numbers to the right.}
\label{sec:box:fig:1d_gamma_von_mu_box}
\end{figure} 

\begin{table}
\caption{\label{sec:box:table1} $\gamma_n(\mu_n)$ for BOX}
\begin{ruledtabular}
\begin{tabular}{ccc}
mode & $\gamma_n(\mu_n=500)$ & $\gamma_n(\mu_n=1000)$ \\
\hline
0 & 436.7686 & 910.5884 \\
1 & 373.5441 & 821.2079 \\
2 & 310.1694 & 731.8231 \\
3 & 245.2187 & 642.2824 \\
4 & 175.4825 & 551.4566 \\
5 & 98.0917 & 457.6764
\end{tabular}
\end{ruledtabular}
\end{table}

\subsection{Gravitational Trap}

Now we solve the GPE with a gravitational potential. With the potential 
\begin{equation}
V_{\rm ext}(x) = 
\begin{cases}
m g x &\text{if } x > 0\\
\infty &\text{else } .
\end{cases}
\end{equation}
we have the natural length scale $L=\left( \hbar^2 / 2 m^2 g\right)^{1/3}$ so that the dimensionless GPE reads
\begin{equation}
\left(- \dfrac{d^2}{dx^2} + x + \gamma \Psi_n(x)^2\right) \Psi_n(x) = \varepsilon_n \Psi_n(x) \, . \label{sec:GOST:1d_gost_eq}
\end{equation}

In the linear case with $\gamma = 0$ the eigenfunctions are well known \cite{springerlink:10.1007/BF00325387}. The general solution is a linear combination of the AiryAi and AiryBi functions where the AiryBi is omitted since it is not compatible with the boundary conditions $\Psi_n(0) = 0$ and $\Psi_n(\infty)=0$.
The $n$-th eigenfunction is given by 
\begin{equation}
\Psi_n(x) = A_n \text{Ai}(x + x_n) \, ,
\end{equation}
where $x_n$ is the $n$-th zero of the AiryAi function and of course the orthogonality relation $\left\langle \Psi_n(x),\Psi_m(x)\right\rangle = \delta_{nm}$ holds.  All zeros are negative so that the normalizable part of the general solution is shifted to the right. The $n$-th eigenvalue $\varepsilon_n$ is also given by the $n$-th zero. Unfortunately no analytic expression for the normalization factor $A_n$ exists. Hence, it is given by
\begin{equation}
\frac{1}{A_n} = \sqrt{\int_{0}^{\infty} dx \text{Ai}(x+x_n)^2}.
\end{equation}

In the nonlinear case it is much more complicated to find solutions because equation (\ref{sec:GOST:1d_gost_eq}) admits a huge number of not normalizable solutions. Most of them have poles on the real axis. 

For $\varepsilon_n = 0$ and $\gamma=2$ problem (\ref{sec:GOST:1d_gost_eq}) is known as the Painlev\'e II equation (PII). This equation possesses the Painlev\'e property \cite{conte2008painleve} which is a condition of integrability. These differential equations cannot be integrated by means of elementary functions. A possibility of finding solutions to PII is to solve the corresponding Riemann-Hilbert problem numerically \cite{springerlink:10.1007/s10208-010-9079-8}. Using this method many different solutions can be found, including nonphysical ones
\footnote{A relatively easy way of solving equation (\ref{sec:GOST:1d_gost_eq}), at least in 1D, is to use the shooting method \cite{Miles01061978}, which is limited by the floating point precision. This is done by integrating the ordinary differential equation numerically from some arbitrary starting point $x_L$ to $x=0$ using an initial guess for  $\Psi_n(x_L),\partial_x\Psi_n(x_L),\varepsilon_n$. 
The initial data can be varied until $\vert\Psi_n(0)\vert < \eta$ for some smallness parameter $\eta$. However, this works very well in 1D problems as long as $\gamma$ is relatively small. With increasing nonlinearity it becomes more and more difficult to find a good guess for the initial data. For example, we were able to find solutions up to $\gamma = 10$.}.

\begin{table}
\caption{\label{sec:gost:table1} $\gamma_{n}(\mu_n)$ for $V_{\rm ext} = x$}
\begin{ruledtabular}
\begin{tabular}{cccc}
mode & $\gamma_{n}(\mu_n=10)$ & $\gamma_{n}(\mu_n=20)$ & $\gamma_{n}(\mu_n=30)$ \\
\hline
0 & 45.4993 & 193.6644 & 442.2494 \\
1 & 36.8594 & 181.1653 & 426.8558 \\
2 & 28.5681 & 168.8259 & 411.5684 \\
3 & 20.6737 & 156.6578 & 396.3996 \\
4 & 13.2327 & 144.6551 & 381.3278 \\
5 & 6.2966 & 132.8302 & 366.3652 \\
6 &  & 121.1916 & 351.5141 \\
7 &  & 109.8536 & 336.4804
\end{tabular}
\end{ruledtabular}
\end{table}

\begin{figure}
\input{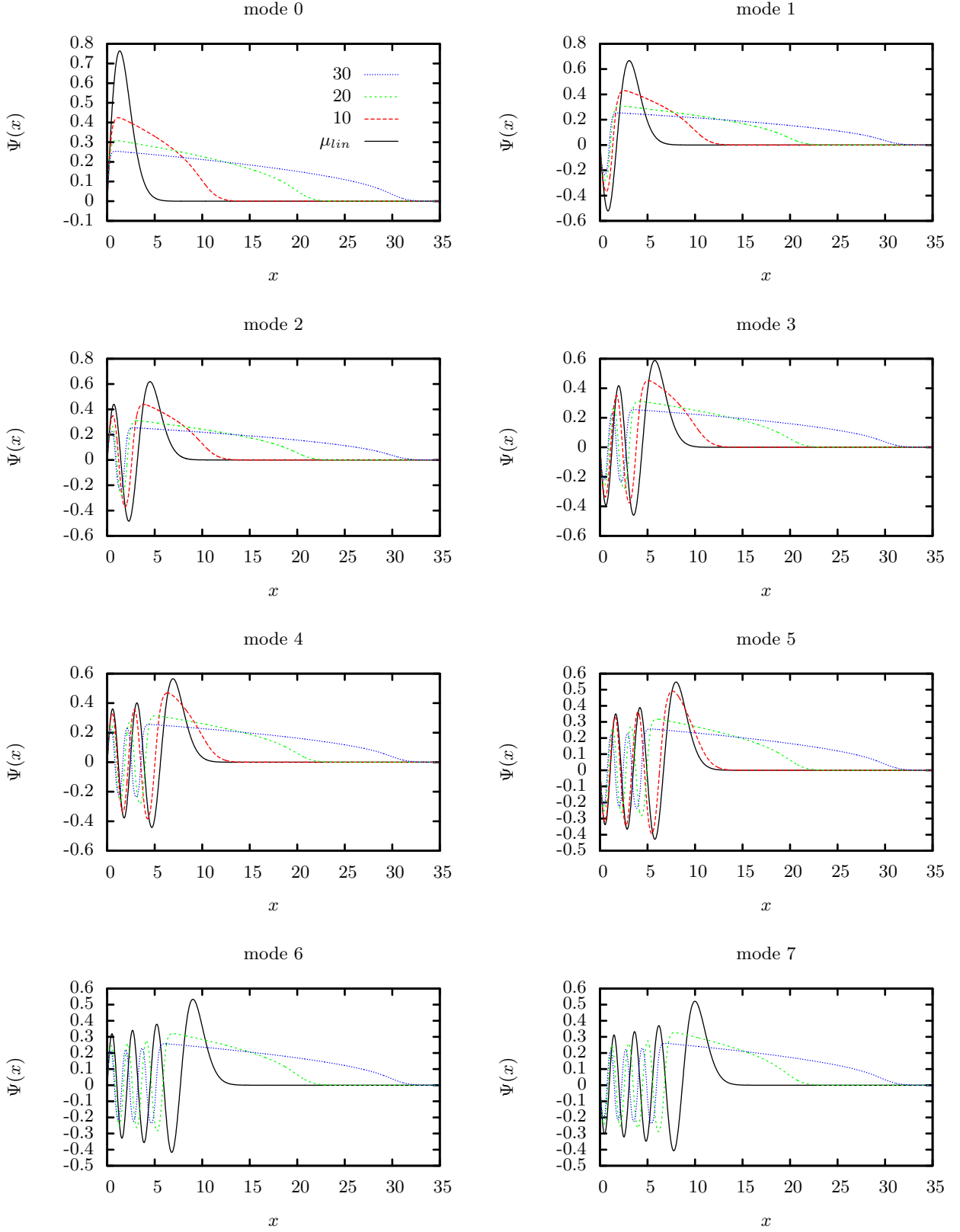}
\caption{The first eight solutions of the GPE for the GOST. The corresponding nonlinearities $\gamma_n$ for given $\mu_n$ can be found in table (\ref{sec:gost:table1}). }
\label{sec:gost:fig:1d_gost}
\end{figure}

In Fig. \ref{sec:gost:fig:1d_gost} the first eight solutions of the GPE with linear potential are calculated for different values of $\mu_n$ with our new method. The solid black lines depict the AiryAi solutions for the linear case. The corresponding nonlinearity factors can be found in table \ref{sec:gost:table1}. The difference here is that the mathematical domain is not finite and that for $x\rightarrow\infty$ the potential is diverging. However, in the numerical implementation the domain has finite size, i.e. of length $L$. On contrary to the standard treatment of boundary conditions it is only necessary to specify the value at $\Psi(0)$ due to the diverging nature of the potential the value at $\Psi(L)$ adjusts itself automatically. We do not pose any boundary conditions explicitly for $\Psi(L)$ so that the finite size domain has no effect on the solutions. 

The first observation is that for a given mode number $n>0$ higher nonlinearities cause a quenching of the region between the boundary and the outermost maximum in comparison with the linear case. This can be understood by taking into account that $V(x=0)=\infty$ limits the space on the left side for encountering particles. As a result, they move towards the outermost maximum causing a depletion of the particle number on the left.

\begin{figure}[h]
\input{1d_gamma_von_mu_gost.tex}
\caption{The one-to-one correspondence between the energy eigenvalues $\mu_s$ and the nonlinearity parameter $\gamma$ for the GOST potential for different eigenmodes. The solid line represents the ground state, increasing mode numbers to the right.}
\label{sec:gost:fig:1d_gamma_von_mu_gost}
\end{figure}

The second observation which seems to be surprising is that the bulk of the wave function for different modes appears not to be changing for fixed $\mu_n$. 
However the explanation is simple: with increasing modes the nonlinearity parameter $\gamma_n$ decreases at fixed $\mu_n$. This behaviour can be clearly seen in Fig. \ref{sec:gost:fig:1d_gamma_von_mu_gost}. Higher nonlinearities always enlarge the bulk of the solution. Fig $\gamma_n(\mu_n)$ in Fig. \ref{sec:gost:fig:1d_gamma_von_mu_gost} shows the one to one correspondence between $\gamma$ and $\mu$.

\subsection{Harmonic Trap}

As a last example we discuss the BEC in a harmonic potential given by 

\begin{equation}
V_{\rm ext} = \frac12 \omega^2 x^2.
\end{equation}

With the natural length scale $L=\sqrt{\hbar/m \omega}$ the dimensionless equation reads
\begin{equation}
\left(-\dfrac{d^2}{dx^2} + x^2  + \gamma \Psi_n(x)^2 \right) \Psi_n(x) = \varepsilon_n \Psi_n(x). \label{sec:TRAP:1d_trap_eq}
\end{equation}

In the linear case $\gamma=0$ the solutions are 
\begin{equation}
\Psi_n(x) = \dfrac{1}{\sqrt{2^n n! \sqrt{\pi}}} \exp\left( -x^2/2 \right) H_n(x) \, ,
\end{equation}
where
\begin{equation}
H_n(x) = \left( -1 \right)^n \exp\left(x^2\right) \frac{d^n}{dx^n} \exp\left(-x^2\right)
\end{equation}
are the weighted Hermite polynomes. 

As far as we know there are no analytic solutions known for this potential for $\gamma \neq 0$. Numerical simulations basically focus on the zeroth mode \cite{PhysRevA.51.1382,PhysRevA.53.2477}. For the zeroth mode there is a rough approximation which can be obtained by neglecting the kinetic energy term in equation (\ref{sec:TRAP:1d_trap_eq}). Then we have an algebraic equation which can easily be solved for $\Psi_0(x)$ and is known as the Thomas--Fermi solution 
\begin{equation}
\Psi_0(x) = \sqrt{\dfrac{\varepsilon_0-x^2}{\gamma}} \, .
\end{equation}
 
\begin{figure}[h]
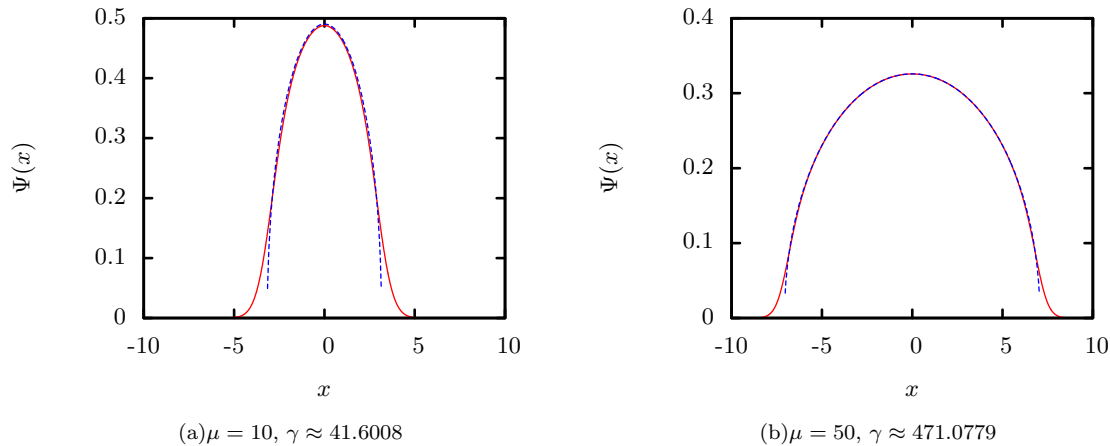

\subfigure[$\mu=10$, $\gamma\approx 41.6008$]{\input{1d_trap_comp_tf_1.tex} }
\subfigure[$\mu=50$, $\gamma\approx 471.0779$]{\input{1d_trap_comp_tf_2.tex} }
\caption{Comparison between the num. solution (solid red line) and the Thomas Fermi approximation (blue dashed line).}
\label{sec:trap:fig:1d_trap_comp_tf}
\end{figure}

In Fig.~\ref{sec:trap:fig:1d_trap_comp_tf} we compared the Thomas--Fermi solution with the numerical solution of the ground state. For large nonlinearities the Thomas--Fermi approximation agrees very well with the numerical results in the center region of the condensate. 

\begin{figure}[h]
\input{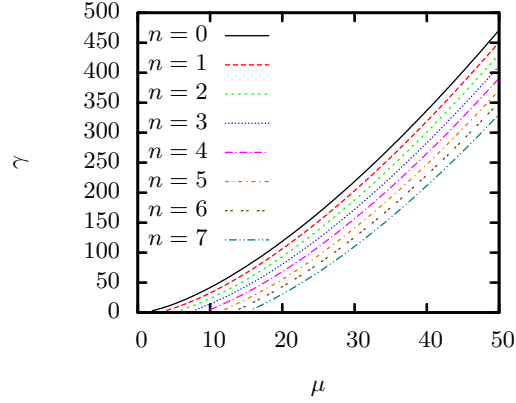}
\caption{The one-to-one correspondence between the energy eigenvalues $\mu_n$ and the nonlinearity parameter $\gamma_n$ for the trapping potential for different eigenmodes. The solid line represents the ground state, increasing mode numbers to the right.}
\label{sec:trap:fig:1d_gamma_von_mu_trap}
\end{figure}

\begin{figure*}
\input{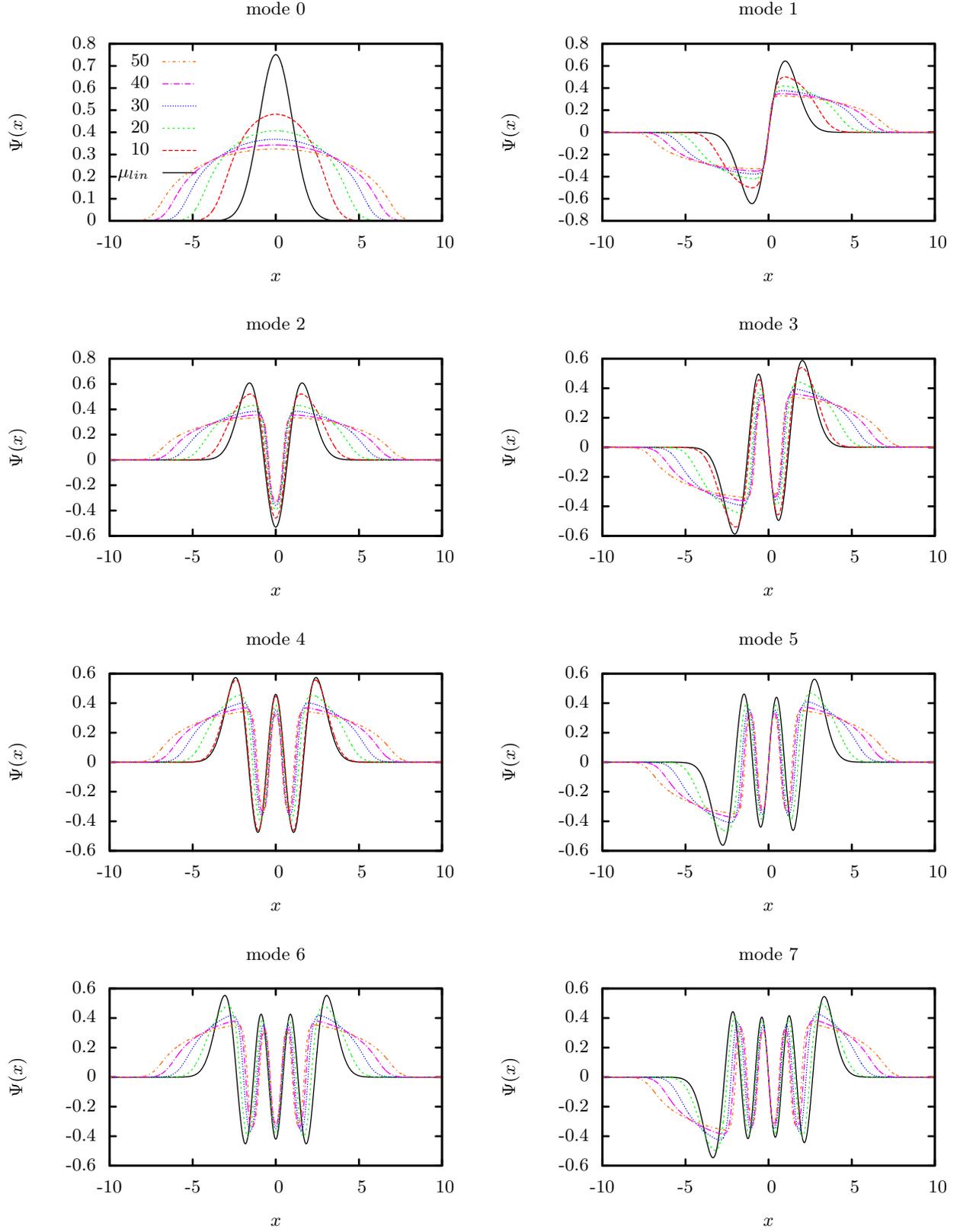}
\caption{The first eight solutions of the GPE with the harmonic trap potential. The corresponding nonlinearities $\gamma_n$ for given $\mu_n$ can be found in table (\ref{sec:trap:table1}). }
\label{sec:trap:fig:1d_trap}
\end{figure*}

In Fig.~\ref{sec:trap:fig:1d_trap} the first eight numerical solutions for the harmonic oscillator potential are depicted for different eigenvalues $\mu_n$ calculated with our new method. The corresponding nonlinearities $\gamma_n$ can be found in table (\ref{sec:trap:table1}). The solid black lines correspond to the linear case with the weighted Hermite polynomial. In the numerical implementation there are no boundary conditions specified. The diverging nature of the potential forces the wave function to decay for $x\rightarrow\infty$. 

The curves in Fig. \ref{sec:trap:fig:1d_trap}  show that with increasing nonlinearity particles from the center region are pushed towards the outer region whereas the inner structures are squeezed. The harmonic potential has a much higher confinement so that the bulk remains relativity small compared to the gravitational potential.

\begin{table}
\caption{\label{sec:trap:table1} $\gamma_n(\mu_n)$ for $V_{\rm ext} = x^2$}
\begin{ruledtabular}
\begin{tabular}{cccccc}
mode & $\gamma_n(\mu_n=10)$ & $\gamma_n(\mu_n=20)$ & $\gamma_n(\mu_n=30)$ & $\gamma_n(\mu_n=40)$ & $\gamma_n(\mu_n=50)$ \\
\hline
0 & 41.6008 & 118.0873 & 218.6985 & 336.9562 & 471.0779\\
1 & 32.5650 & 106.1331 & 203.1434 & 319.0604 & 451.0741\\
2 & 23.4540 & 93.4226  & 187.6607 & 301.1403 & 431.0489\\
3 & 14.2610 & 80.7247  & 172.1164 & 283.2037 & 411.0061\\
4 & 4.8558  & 68.0688  & 156.5821 & 265.2629 & 390.9532\\
5 &         & 55.5007  & 141.0820 & 247.3326 & 370.8997\\
6 &         & 43.0556  & 125.6424 & 229.4285 & 350.8563\\
7 &         & 30.7438  & 110.2917 & 211.5678 & 330.8343
\end{tabular}
\end{ruledtabular}
\end{table}

\section{Discussion and outlook}

In this article we presented a new algorithm that is capable to find higher Morse index solutions of the stationary GPE for large nonlinearities. Mathematically speaking, these are saddle point solutions. The three crucial points are (i) to start with a fixed eigenvalue $\mu$, (ii) the reference point that contains only a function of the same Morse index in the support and (iii) the preconditioning of the $L_2$ gradient by using the analytic expression for the Hessian.
Furthermore we demonstrated that we can find solutions for the GPE with large nonlinearity parameter in external potentials by starting only with the eigenfunction for the n-th mode of the corresponding linear problem.

In summary we calculated the eigenfunctions and energies for the one dimensional GPE for three classical potentials: the box, the harmonic trap and the GOST. In the case of the GOST we obtained higher order modes up to order seven for large nonlinearity parameters in the range of $\gamma= 336-442$. To our present knowledge this seems to be the first time that solutions to the GOST setup for such high nonlinearities and high modes have been calculated. Furthermore, we showed that in the case of the box the numerically found solutions completely agree with known analytical solutions which confirms our algorithm. Also, there is a good agreement between the numerical zero mode solutions for the trap and the Thomas-Fermi approximation in the central region of the BEC.

The next logical step is to apply this algorithm in $2D$ and $3D$ setups of the aforementioned three cases so that more realistic physical systems will be modelled. 
Moreover, the physical stability may be checked by propagating the solutions in time. 
Another issue which may be treated in future is to include self--gravity effects. At first, for an efective equation as the GPE self--gravity should be considered. For very dilute gases one may expect no effects but for high density BECs corresponding effects should be estimated. Self gravity also is an idea stated by Penrose \cite{Penrose95} to understand the collapse of the wave function. Therefore it might be of interest to investigate whether in this context such effects might be accessible to experiment. We also plan to adopt our method to the case of coupled many component GPEs. 

Finally, concerning the algorithm, in the future it may be interesting to find a new stepsize control that incorporates the ascent direction. This may improve convergence behaviour. For spatial dimensions larger than 1 we also would like to extend our algorithm in order to incorporate also different coordinate systems which are more adapted to the physical problem. 

\section*{Acknowledgement}

We would like to thank S. Herrmann and V. Perlick for many fruitful discussions. E.G. and \v Z.M. gratefully acknowledge financial support from DLR, project number 50WM-0942 C.L. also thanks the cluster of excellence QUEST for support. 

\appendix

\section{Search direction}

For the action $A$ we can write down the Taylor expansion around $\phi^k$ up to first order:

\begin{equation}
A\left[ \phi^{k+1}; \mu_s \right] = A\left[ \phi^{k}; \mu_s \right] + A^{\prime}\left[ \phi^{k}; \mu_s \right] \left( \phi^{k+1} - \phi^k \right) + \mathcal{O} \left( \left( \phi^{k+1} - \phi^k \right)^2 \right) \label{sec:apd:1}.
\end{equation}

Using the Newton step (\ref{sec:alg:newton_method}) and the search direction (\ref{sec:alg:myprecond}) equation (\ref{sec:apd:1}) can be written as:

\begin{align}
A\left[ \phi^{k+1}; \mu_s \right] &= A\left[ \phi^{k}; \mu_s \right] - \tau A^{\prime}\left[ \phi^{k}; \mu_s \right] d^k + \mathcal{O} \left( d^2 \right) \nonumber\\
 &= A\left[ \phi^{k}; \mu_s \right] - \tau \langle \nabla_{L_2}\mathcal{A}\left[\phi^k;\mu_s\right], d^k\rangle + \mathcal{O} \left( d^2 \right) .
\end{align}

If $A\left[ \phi^{k+1}; \mu_s \right] < A\left[ \phi^{k}; \mu_s \right]$ and $\langle \nabla_{L_2}\mathcal{A}\left[\phi^k;\mu_s\right], d^k\rangle>0$ then $-d^k$ is a descent direction. If $A\left[ \phi^{k+1}; \mu_s \right] > A\left[ \phi^{k}; \mu_s \right]$ and $\langle \nabla_{L_2}\mathcal{A}\left[\phi^k;\mu_s\right], d^k\rangle<0$ then $-d^k$ is a ascent direction.

\bibliography{paper}

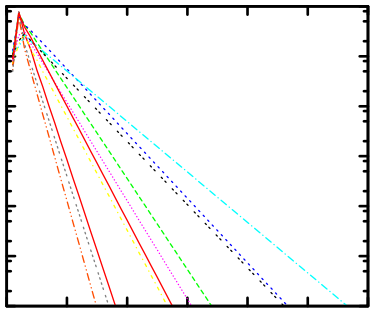
\end{document}